\documentclass[11pt,a4paper,french,english,PSfrag,BCOR8mm, most, usenames, dvipsnames]{article}
\usepackage[T1]{fontenc}
\usepackage[latin9]{inputenc}
\usepackage{color}
\usepackage{babel}
\usepackage{tcolorbox}
\usepackage{amsmath}
\usepackage{amsthm}
\usepackage{amssymb}
\usepackage{graphicx}
\usepackage[pdfusetitle,
 bookmarks=true,bookmarksnumbered=false,bookmarksopen=false,
 breaklinks=false,pdfborder={0 0 0},pdfborderstyle={},backref=false,colorlinks=true]
 {hyperref}

\makeatletter

\pdfpageheight\paperheight
\pdfpagewidth\paperwidth

\providecommand{\tabularnewline}{\\}

\numberwithin{equation}{section}
\numberwithin{figure}{section}
\theoremstyle{plain}
\newtheorem{thm}{\protect\theoremname}[section]
\theoremstyle{definition}
\newtheorem{defn}[thm]{\protect\definitionname}
\theoremstyle{remark}
\newtheorem{rem}[thm]{\protect\remarkname}
\theoremstyle{plain}
\newtheorem{prop}[thm]{\protect\propositionname}

\usepackage[a4paper, total={17cm, 27cm}]{geometry}
\usepackage{color}
\usepackage{graphics}
\usepackage{stackrel} 

\usepackage{stmaryrd}

\makeatother

\addto\captionsenglish{\renewcommand{\definitionname}{Definition}}
\addto\captionsenglish{\renewcommand{\propositionname}{Proposition}}
\addto\captionsenglish{\renewcommand{\remarkname}{Remark}}
\addto\captionsenglish{\renewcommand{\theoremname}{Theorem}}
\addto\captionsfrench{\renewcommand{\definitionname}{Définition}}
\addto\captionsfrench{\renewcommand{\propositionname}{Proposition}}
\addto\captionsfrench{\renewcommand{\remarkname}{Remarque}}
\addto\captionsfrench{\renewcommand{\theoremname}{Théorème}}
\providecommand{\definitionname}{Definition}
\providecommand{\propositionname}{Proposition}
\providecommand{\remarkname}{Remark}
\providecommand{\theoremname}{Theorem}

\begin{document}
\selectlanguage{english}
\title{Topological index formula in physical waves: spectral flow, Chern
index and topological contacts}
\author{\href{mailto:leonmonnier1@gmail.com}{L\' eon Monnier}\\
 {\small Univ. Grenoble Alpes, CNRS, Institut Fourier, F-38000 Grenoble,
France}\\
 {\small\href{mailto:leonmonnier1@gmail.com}{leonmonnier1@gmail.com}}
\\
\and\\
 \href{https://www-fourier.ujf-grenoble.fr/~faure/}{Fr\' ed\' eric Faure}\\
 {\small Univ. Grenoble Alpes, CNRS, Institut Fourier, F-38000 Grenoble,
France}\\
 {\small\href{mailto:frederic.faure@univ-grenoble-alpes.fr}{frederic.faure@univ-grenoble-alpes.fr}} }
\date{2025 july 11}
\maketitle
\begin{abstract}
We study a family of pseudodifferential operators (quantum Hamiltonians)
on $L^{2}(\mathbb{R}^{n};\mathbb{C}^{d})$ whose spectrum exhibits
two energy bands exchanging a finite number of eigenvalues. We show
that this number coincides with the Chern index of a vector bundle
associated to the principal symbol (the classical Hamiltonian). This
result provides a simple yet illustrative instance of the Atiyah-Singer
index formula, with applications in areas such as molecular physics,
plasma physics or geophysics. We also discuss the phenomenon of topological
contact without exchange between energy bands -- a feature that cannot
be detected by the Chern index or K-theory, but rather reflects subtle
torsion effects in the homotopy groups of spheres.

\footnote{

81V55 Molecular physics,

55R50 Stable classes of vector space bundles, K-theory,

47A53 (Semi-) Fredholm operators; index theories,

19K56 Index theory,

81Q20 Semiclassical techniques, including WKB and Maslov methods

81Q05 Closed and approximate solutions to the Schrödinger, Dirac,
Klein-Gordon and other equations of quantum mechanics}
\end{abstract}
\newpage{}

\tableofcontents{}
\newtcolorbox{cBoxA}[1][]{enhanced, frame style={purple!80}, interior style={red!0}, #1}

\newtcolorbox{cBoxB}[2][]{enhanced, frame style={teal!80}, interior style={cyan!0}, #2}

\global\long\def\eq#1{\underset{(#1)}{=}}%

\section{Introduction}

Topological phenomena refer to properties that remain invariant under
continuous deformations of a model. These properties often lead to
particularly robust physical behaviors, which are preserved under
perturbations---for example, in the context of the quantum Hall effect.
To study such models mathematically, tools from algebraic topology
are essential. In particular, the Atiyah--Singer index theorem plays
a central role by connecting topological invariants with operator
theory and spectral properties.

In this paper, we present and prove a particular case of this theorem,
following the approach of \cite[Arxiv version]{faure_manifestation_topol_index_2019}.
We consider a family of operators indexed by a parameter $\mu$, whose
spectrum consists of two energy bands (i.e. group of levels) separated
by a spectral gap. As $\mu$ varies, a finite number of eigenvalues
are exchanged between the bands. Using microlocal analysis, we derive
an index formula (\ref{eq:formule_indice}) in theorem \ref{thm:Index-formula-Let}
that relates this number $\mathcal{N}$ to the Chern index $\mathcal{C}$
of a vector bundle constructed from the symbol of the operator, in
a simple way since we get that $\mathcal{N}=\mathcal{C}$.

This general framework has applications in geophysical fluid dynamics---for
instance, in the study of equatorial waves \cite{matsuno1966quasi}\cite{Delplace_Venaille_2018}---as
well as in plasma physics \cite{2022_qin_plasma_physics}. In the
remainder of the article, we first introduce a simple family of operators
exhibiting the key topological features, and then use this setting
to establish a more general index formula.

\section{\protect\label{sec:Normal-form-model}Normal form model in dimension
$n=1$}

We will use $\hat{x},\hat{p}$ to be the usual position and momentum
operators acting on $\psi\in L^{2}\left(\mathbb{R}\right)$ defined
by $\left(\hat{x}\psi\right)(x)=x\psi(x)$ and $\left(\hat{p}\psi\right)(x)=-i\frac{\partial\psi}{\partial x}(x)$
and let
\begin{equation}
a:=\frac{1}{\sqrt{2}}\left(\hat{x}+i\hat{p}\right),\quad a^{\dagger}:=\frac{1}{\sqrt{2}}\left(\hat{x}-i\hat{p}\right).\label{eq:a_a+}
\end{equation}

\begin{cBoxA}{}
\begin{defn}
We consider the following family of operators $\hat{E}_{\mu}$ indexed
by a real parameter $\mu\in\mathbb{R}$, acting in the Hilbert space
$L^{2}(\mathbb{R})\otimes\mathbb{C}^{2}$.
\end{defn}

\begin{equation}
\hat{E}_{\mu}:=\left(\begin{array}{cc}
-\mu & \hat{x}+i\hat{p}\\
\hat{x}-i\hat{p} & +\mu
\end{array}\right)\eq{\ref{eq:a_a+}}\left(\begin{array}{cc}
-\mu & \sqrt{2}\,a\\
\sqrt{2}\,a^{\dagger} & \mu
\end{array}\right)\label{eq:symbole_H_mu-1}
\end{equation}
\end{cBoxA}

Notice that $\hat{E}_{\mu}$ is self-adjoint.
\begin{rem}
This normal form model appears in various areas of physics. For example,
in \cite{fred-boris,fred-boris01}, it is shown how this normal form
provides a microlocal description of the interaction between the fast
vibrational motion and the slow rotational motion of the molecule
depicted in Figure \ref{fig:Niveaux-d'=00003D0000E9nergie-de}. In
short, $\left(\hat{x},\hat{p}\right)$ represent the quantization
of local coordinates on the sphere $S^{2}$ , which describes the
rotational motion of the molecule, while the $\mathbb{C}^{2}$ space
encodes the quantum dynamics of the fast vibrations of the molecule,
restricted to an effective two-level system.

\begin{figure}
\begin{centering}
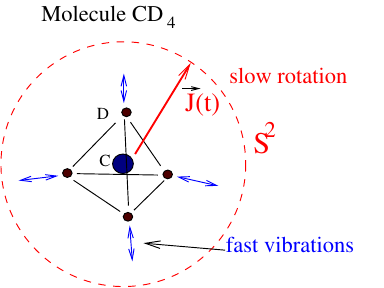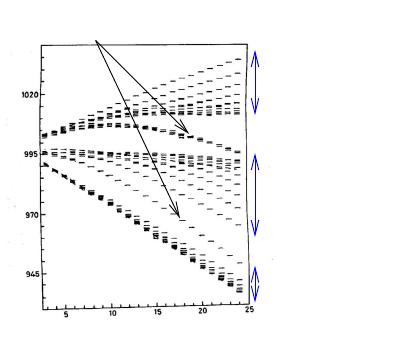
\par\end{centering}
\caption{\protect\label{fig:Niveaux-d'=00003D0000E9nergie-de}Energy levels
(in $cm^{-1}$) of the $CD_{4}$ molecule (carbon with four deuterium
atoms) as a function of the total angular momentum $J\in\mathbb{N}$
(a conserved quantity corresponding to rotational energy). The fine
structure of the spectrum reflects the slow rotational motion, while
the broad structure corresponds to the faster vibrational dynamics.
The spectrum exhibits clusters of energy levels, with some levels
crossing or connecting different clusters \cite{fred-boris,fred-boris01,fred-boris02,fred-mikael-04,boris1}.}
\end{figure}
\end{rem}

\subsection{Spectral flow and spectral index}

\begin{figure}
\begin{centering}
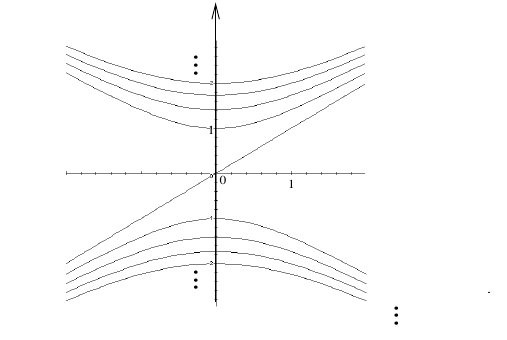
\par\end{centering}
\caption{\protect\label{fig:Spectre-de-().}Spectrum of (\ref{eq:modele}).}
\end{figure}

The next proposition describes the spectrum of the operator defined
in Equation (\ref{eq:symbole_H_mu}) and exibits a single eigenvalue
that moves upward, see in Figure \ref{fig:Spectre-de-().}. This behavior
defines a spectral index $\mathcal{N}_{E}=+1$. Throughout this paper,
this elementary model will be referred to as model $E$.

\begin{cBoxB}{}
\begin{prop}
\textbf{\label{prop:Spectre-de-.}<<Spectrum of $\hat{E}_{\mu}$>>}.
For each parameter $\mu\in\mathbb{R}$, the operator $\hat{E}_{\mu}$,
(\ref{eq:symbole_H_mu}), has discrete spectrum in $L^{2}\left(\mathbb{R}_{x}\right)\otimes\mathbb{C}^{2}$
given by 
\begin{equation}
\hat{E}_{\mu}\phi_{n}^{\pm}=\omega_{n}^{\pm}\phi_{n}^{\pm},\quad n\geq1,\label{eq:modele}
\end{equation}
with for any $n\in\mathbb{N}\backslash\left\{ 0\right\} $, eigenvalues
and eigenvectors are
\begin{align}
\omega_{n}^{\pm} & =\pm\sqrt{\mu^{2}+2n}\label{eq:spectre}\\
\phi_{n}^{\pm} & =\left(\begin{array}{c}
\frac{\sqrt{2n}}{\mu+\omega_{n}^{\pm}}\varphi_{n-1}\\
\varphi_{n}
\end{array}\right)\nonumber 
\end{align}
and additionally for $n=0$, 
\[
\hat{E}_{\mu}\phi_{0}=\omega_{0}\phi_{0},
\]
with 
\[
\omega_{0}=\mu,\qquad\phi_{0}=\left(\begin{array}{c}
0\\
\varphi_{0}
\end{array}\right).
\]
$\left(\varphi_{n}\right)_{n\geq0}$ are the Hermite functions in
$L^{2}\left(\mathbb{R}_{x}\right)$ of the harmonic oscillator defined
by 
\begin{equation}
\varphi_{0}\left(x\right)=\frac{1}{\pi^{1/4}}e^{-\frac{1}{2}x^{2}},\quad\varphi_{n+1}=\frac{1}{\sqrt{n+1}}a^{\dagger}\varphi_{n},\qquad a\varphi_{n}=\sqrt{n}\varphi_{n-1}.\label{eq:phi_0}
\end{equation}
\end{prop}

\end{cBoxB}

\begin{proof}
By direct computation, see \cite[Arxiv version]{faure_manifestation_topol_index_2019}
for details.
\end{proof}

See figure \ref{fig:Spectre-de-().}. We remark that for $|\mu|>1$
the spectrum of $\hat{E}_{\mu}$ has no eigenvalues in the interval
$\left[-1,1\right]$, this is called a spectral gap. When $\mu$ goes
from $-1$ to $1$, we observe that there is $\mathcal{N}_{E}=1$
eigenvalue passing through this gap. Below, in a more general model,
we will see that $\mathcal{N}_{E}\in\mathbb{Z}$ is a \textbf{spectral
index}. To study the origin of this phenomenon, we remark that the
operator $\hat{E}_{\mu}$ can be seen as the quantization\footnote{Here quantization means that the position/momentum variables $x,p$
are replaced by the operators $\hat{x},\hat{p}$. We will give a more
precise definition later.} of the following matrix valued function $E_{\mu}:\mathbb{R}^{2}\rightarrow\mathrm{Herm}\left(\mathbb{C}^{2}\right)$
on a phase space.

\begin{cBoxA}{}
\begin{defn}
The matrix valued function on a phase space.
\begin{equation}
E_{\mu}:\mathbb{R}^{2}\rightarrow E_{\mu}\left(x,p\right):=\left(\begin{array}{cc}
-\mu & x+ip\\
x-ip & +\mu
\end{array}\right)\in\mathrm{Herm}\left(\mathbb{C}^{2}\right).\label{eq:symbole_H_mu}
\end{equation}
is called the \textbf{symbol} of $\hat{E}_{\mu}$ (also called classical
Hamiltonian function in physics).
\end{defn}

\end{cBoxA}

\subsection{Chern index}

In the next proposition, for each parameter $\left(\mu,x,p\right)\in\mathbb{R}^{3}\backslash\left\{ 0\right\} $,
we will consider the lower eigenvector $\psi\left(\mu,x,p\right)\in\mathbb{C}^{2}$
of the $2\times2$ matrix $E_{\mu}\left(x,p\right)$ in (\ref{eq:symbole_H_mu}).
Recall that this eigenvector is defined only up to a scalar, hence
only the (complex) one dimensional eigenspace $F_{-}\left(\mu,x,p\right)=\left\{ \lambda\psi\left(\mu,x,p\right),\lambda\in\mathbb{C}\right\} \subset\mathbb{C}^{2}$
is well defined. We get a family of one dimensional vector spaces
$F_{-}:=\left\{ F_{-}\left(\mu,x,p\right)\subset\mathbb{C}^{2},\left(\mu,x,p\right)\in\mathbb{R}^{3}\backslash\left\{ 0\right\} \right\} $.
This family $F_{-}$ is called a \textbf{complex vector bundle of
rank $1$ over the space of parameters $\mathbb{R}^{3}\backslash\left\{ 0\right\} $.
}We are interested by the ``topology'' (or isomorphism class)\footnote{meaning this family up to equivalence under continuous (or homotopic)
bundle map.} of this vector bundle $F_{-}$.

This topology is characterized by an integer $\mathcal{C}\left(F_{-}\right)\in\mathbb{Z}$
called the \textbf{Chern index}. There are many ways to define and
compute it. One of the simplest is the following.

\begin{cBoxA}{}
\begin{defn}
\label{def:-We-consider}\cite{hatcher_ktheory} We consider the unit
sphere $S^{2}=\left\{ \left(\mu,x,p\right)\in\mathbb{R}^{3},\left\Vert \left(\mu,x,p\right)\right\Vert =1\right\} $
that can be decomposed as the union of two hemispheres joined at the
equator $S^{1}$. On each hemisphere separately $H_{1}$ (respect.
$H_{2}$), one can choose continuously an eigenvector $s_{1}$ (respect.
$s_{2}$) in $F_{-}$ (but not globally on $S^{2}$). Then one observe
that on each point $\theta\in S^{1}$ of the equator, $s_{1},s_{2}$
are related by a phase $s_{2}=e^{i\varphi\left(\theta\right)}s_{1}$,
giving a map $\varphi:\theta\in S_{1}\mapsto\varphi\left(\theta\right)\in S^{1}$,
called the \textbf{clutching function}. The \textbf{Chern index} is
the \href{https://en.wikipedia.org/wiki/Winding_number}{winding number}
(or \href{https://en.wikipedia.org/wiki/Degree_of_a_continuous_mapping}{degree},
see (\ref{eq:def_degre})) of this map:
\[
\mathcal{C}\left(F_{-}\right):=\mathrm{deg}\left(\varphi\right)\in\mathbb{Z}.
\]
\end{defn}

\end{cBoxA}

\begin{figure}

\begin{centering}
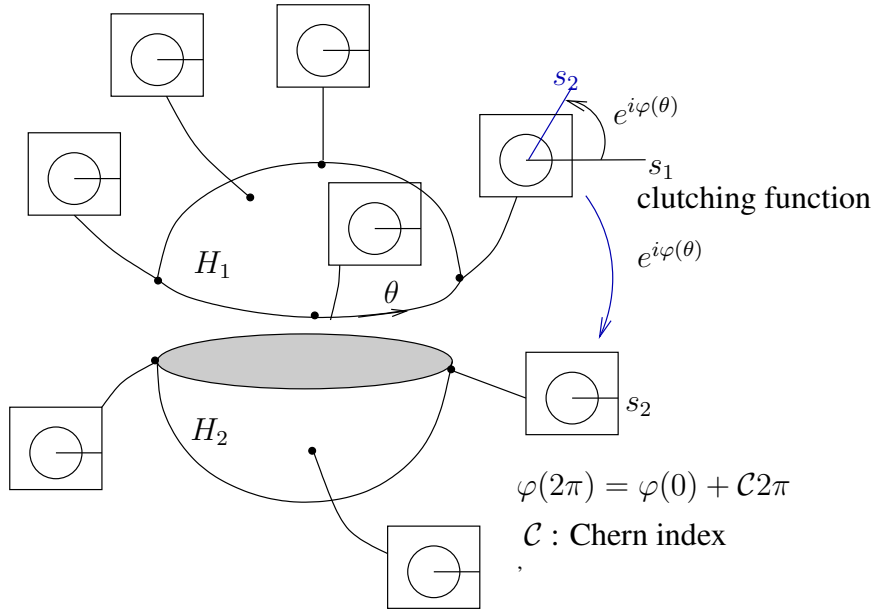\caption{\protect\label{fig:Chern-index-}Chern index $\mathcal{C}$ of a rank
one bundle over $S^{2}$, computed by the winding number of the clutching
function.}
\par\end{centering}
\end{figure}

The proof written below gives more details. See also figures \ref{fig:Chern-index-}
and \ref{fig:We-have-,}.

\begin{figure}
\begin{centering}
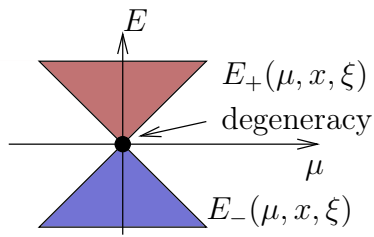
\par\end{centering}
\caption{\protect\label{fig:We-have-,}Eigenvalues from (\ref{eq:val_p}).
We have $\omega^{-}\left(\mu,x,p\right)\protect\leq-\left|\mu\right|$,
$\omega^{+}\left(\mu,x,p\right)\protect\geq\left|\mu\right|$. The
red domain represents the possible values of $\omega_{-}\left(\mu,x,p\right)$
with $\mu$ fixed and $\left(x,p\right)\in\mathbb{R}^{2}$. Similarly,
the blue domain represent $\omega_{-}\left(\mu,x,p\right)$. The degeneracy
is at $\left(\mu,x,p\right)=\left(0,0,0\right)$.}
\end{figure}

\begin{cBoxB}{}
\begin{prop}
\textbf{\label{prop:Aspects-topologiques-du-1}<<Topological aspects
of the symbol $E_{\mu}$>>}. The eigenvalues of the matrix $E_{\mu}\left(x,p\right)\in\mathrm{Herm}\left(\mathbb{C}^{2}\right)$
, Eq.(\ref{eq:symbole_H_mu}), are
\begin{equation}
\omega_{\pm}\left(\mu,x,p\right)=\pm\sqrt{\mu^{2}+x^{2}+p^{2}}\label{eq:val_p}
\end{equation}
There is therefore a degeneracy $\omega_{+}=\omega_{-}$ (only) for
$\left(\mu,x,p\right)=\left(0,0,0\right)$. To each point on the unit
sphere $\left(\mu,x,p\right)\in S^{2}=\left\{ \left(\mu,x,p\right)\in\mathbb{R}^{3},\left\Vert \left(\mu,x,p\right)\right\Vert =1\right\} $,
we can associate the eigenspace $F_{-}\left(\mu,x,p\right)\subset\mathbb{C}^{2}$
corresponding to the lower eigenvalue $\omega_{-}(\mu,x,p)$. The
Chern index of this rank $1$ vector bundle $F_{-}$ is:
\[
\mathcal{C}_{E}:=\mathcal{C}\left(F_{-}\right)=+1.
\]
\end{prop}

\end{cBoxB}

\begin{proof}
See \cite[Arxiv version]{faure_manifestation_topol_index_2019} for
other alternative proofs. We have 
\[
q\left(\omega\right):=\mathrm{det}\left(\omega\mathrm{Id}-E_{\mu}\left(x,p\right)\right)=\mathrm{det}\left(\begin{array}{cc}
\omega+\mu & -\left(x+ip\right)\\
-\left(x-ip\right) & \omega-\mu
\end{array}\right)=\omega^{2}-\left(\mu^{2}+x^{2}+p^{2}\right)
\]
hence $q\left(\omega\right)=0$ gives eigenvalues $\omega_{\pm}=\pm r$
with $r:=\sqrt{\mu^{2}+x^{2}+p^{2}}$, i.e. Eq. (\ref{eq:val_p}).
The eigenvectors of $E_{\mu}$ are respectively\footnote{In \href{https://www-fourier.ujf-grenoble.fr/~parisse/xcasfr.html}{xcas online},
write: \texttt{E:={[}{[}-mu,x+i{*}xi{]},{[}x-i{*}xi,mu{]}{]}; eigenvals(E);
eigenvects(E);}}
\begin{equation}
U_{+}=\left(\begin{array}{c}
-\mu+r\\
x-ip
\end{array}\right),\quad U_{-}=\left(\begin{array}{c}
-\mu-r\\
x-ip
\end{array}\right),\label{eq:U_+_U_-}
\end{equation}
i.e. $E_{\mu}\left(x,p\right)U_{\pm}=\omega_{\pm}U_{\pm}$. Write
$F_{\pm}\left(\mu,x,p\right):=\mathrm{Vect}\left(U_{\pm}\right)\subset\mathbb{C}^{2}$
the associated eigenspaces. The spectral projector $\pi_{-}$ on $F_{-}$
is
\begin{equation}
\pi_{-}=\frac{1}{\left\Vert U_{-}\right\Vert ^{2}}U_{-}\langle U_{-}|.\rangle\quad:\mathbb{C}^{2}\rightarrow F_{-}\left(\mu,x,p\right).\label{eq:pi_-}
\end{equation}
Consider $S^{2}=\left\{ \left(\mu,x,p\right)\in\mathbb{R}^{3},r=\left\Vert \left(\mu,x,p\right)\right\Vert =1\right\} $
the unit sphere in the parameter space and the northern and southern
hemispheres $H_{1}:=\left\{ \left(\mu,x,p\right)\in S^{2},\mu\geq0\right\} $,
$H_{2}:=\left\{ \left(\mu,x,p\right)\in S^{2},\mu\leq0\right\} $.
The projection of the fixed vector $\left(\begin{array}{c}
1\\
0
\end{array}\right)\in\mathbb{C}^{2}$ on $F_{-}$ gives the global \href{https://en.wikipedia.org/wiki/Vector_bundle\#Sections_and_locally_free_sheaves}{section}:
\begin{equation}
s_{1}\left(\mu,x,p\right):=\pi_{-}\left(\begin{array}{c}
1\\
0
\end{array}\right)\eq{\ref{eq:pi_-}}\frac{\left(-\mu-1\right)}{\left(\left(\mu+1\right)^{2}+x^{2}+p^{2}\right)}\left(\begin{array}{c}
-\mu-1\\
x-ip
\end{array}\right).\label{eq:s1}
\end{equation}
We have $\left\Vert s_{1}\right\Vert ^{2}=\frac{\left(\mu+1\right)^{2}}{\left(\left(\mu+1\right)^{2}+x^{2}+p^{2}\right)}=\frac{1+\mu}{2}$
hence $\left\Vert s_{1}\right\Vert ^{2}\neq0$ on $H_{1}$. Hence
$s_{1}$ is a trivialization of $F_{-}\rightarrow H_{1}$ (i.e. a
non zero section). We consider also the following trivialization of
$F_{-}\rightarrow H_{2}$:
\begin{equation}
s_{2}\left(\mu,x,p\right):=\pi_{-}\left(\begin{array}{c}
0\\
1
\end{array}\right)=\frac{\left(x+ip\right)}{\left(\left(\mu+1\right)^{2}+x^{2}+p^{2}\right)}\left(\begin{array}{c}
-\mu-1\\
x-ip
\end{array}\right),\label{eq:s2}
\end{equation}
We have $\left\Vert s_{2}\right\Vert ^{2}=\frac{\left(x^{2}+p^{2}\right)}{\left(\left(\mu+1\right)^{2}+x^{2}+p^{2}\right)}=\frac{1-\mu}{2}$
hence $\left\Vert s_{2}\right\Vert ^{2}\neq0$ on $H_{2}$. The clutching
function on the equator $S^{1}=\left\{ \mu=0,x+ip=e^{i\theta},\theta\in[0,2\pi[\right\} $
is defined by 
\begin{align*}
s_{2}\left(\theta\right) & =f_{21}\left(\theta\right)s_{1}\left(\theta\right)\\
\Leftrightarrow & \left(x+i\xi\right)\left(\begin{array}{c}
-1\\
x-ip
\end{array}\right)=-f_{21}\left(\theta\right)\left(\begin{array}{c}
-1\\
x-ip
\end{array}\right)\\
\Leftrightarrow & f_{21}\left(\theta\right)=-e^{i\theta}.
\end{align*}
The \href{https://en.wikipedia.org/wiki/Degree_of_a_continuous_mapping}{degree}
(or winding number) of the function $f_{21}:\theta\in S^{1}\rightarrow f_{21}\left(\theta\right)=-e^{i\theta}\in U\left(1\right)\equiv S^{1}$
is $\mathcal{C}=\mathrm{deg}\left(f_{21}\right)=+1$.
\end{proof}

\subsection{Conclusion}

In this very simple normal form model $E$, we have observed that
there is $\mathcal{N}_{E}=1$ eigenvalue in the spectral flow of the
quantum model $\left(\hat{E}_{\mu}\right)_{\mu\in\mathbb{R}}$ and
a Chern index $\mathcal{C}_{E}:=\mathcal{C}\left(F_{-}\right)=1$
obtained from the eigenspaces of the symbol functions $\left(E_{\mu}\right)_{\mu\in\mathbb{R}}$.
We also remark that the following equality holds true 
\begin{equation}
\mathcal{C}_{E}=\mathcal{N}_{E}=1.\label{eq:index_formula_E}
\end{equation}
One may wonder whether this equality is merely a coincidence. We will
explain that there is a very general theory called \href{https://en.wikipedia.org/wiki/Atiyah\%E2\%80\%93Singer_index_theorem}{\textquotedblleft Index theory\textquotedblright}
\cite{booss_85} which accounts for this relation called ``index
formula'' and holds in a much more general setting (arbitrary dimensions
and nonlinear symbols in $\left(x,p\right)$). For this, we need microlocal
analysis. We present this more general model below.

\section{General model}

Here, we generalize the model introduced in Section \ref{sec:Normal-form-model}.
The general setting is defined in terms of a symbol (a matrix-valued
function on phase space) under specific hypotheses that ensure both
the existence of a spectral index\emph{ }$\mathcal{N}$ for the quantized
operators and a topological Chern index $\mathcal{C}$ associated
with a vector bundle constructed from the symbol. We then prove the
equality $\mathcal{C}=\mathcal{N}$.

\subsection{Definitions}

\subsubsection{Hypothesis on the symbol}

We first define the model by its symbol: a matrix valued function
on phase space depending on a parameter $\mu$. Below, $\mathrm{Herm(\mathbb{C}^{d})}$
denotes the vector space of hermitian matrices of dimension $d\times d$.
There are (important but usual) conditions at infinity on the symbol
that we will ignore in this article \cite{grigis_sjostrand}.

\begin{cBoxA}{}
\begin{defn}
\cite[ass. 2.1]{faure_manifestation_topol_index_2019}\label{assump}Our
model $H$ is defined from a continuous family of symbols depending
on a parameter $\mu\in\mathbb{R}$:
\begin{equation}
H_{\mu}:\left(x,p\right)\in\mathbb{R}^{2n}\to H_{\mu}\left(x,p\right)\in\mathrm{Herm}(\mathbb{C}^{d})\label{eq:symbol_H_miu}
\end{equation}

such that, if we denote the real and sorted eigenvalues of the matrix
$H_{\mu}\left(x,p\right)$ by 
\begin{equation}
\omega_{1}\left(\mu,x,p\right)\leq\ldots\leq\omega_{d}\left(\mu,x,p\right),\label{eq:eigenvalues}
\end{equation}
we suppose that there exists an index $r\in\left\{ 1,\ldots d-1\right\} $
and $C>0$ such that for every $\left(\mu,x,p\right)\in\mathbb{R}^{1+2n}$
with $\left\Vert \left(\mu,x,p\right)\right\Vert \geq1$, and $\mu\in\left(-2,2\right)$,
we have the \textbf{gap assumption}
\begin{equation}
\omega_{r}\left(\mu,x,p\right)<-C\text{ and }\omega_{r+1}\left(\mu,x,p\right)>+C.\label{eq:gap}
\end{equation}
\end{defn}

\end{cBoxA}

\begin{figure}
\begin{centering}
\scalebox{0.9}[0.9]{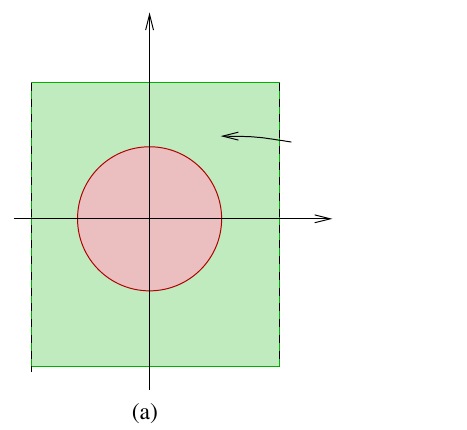}~~~~~\scalebox{0.9}[0.9]{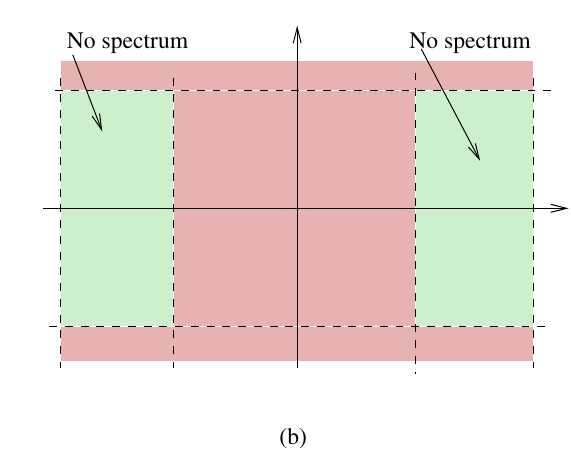}
\par\end{centering}
\caption{\protect\label{fig:Hypoth=00003D0000E8se}Illustration of the assumption
(\ref{eq:gap}). On figure (a), for parameters $\left(\mu,x,p\right)\in\mathbb{R}\times\mathbb{R}^{n}\times\mathbb{R}^{n}$
in the green domain, we assume that the spectrum of the hermitian
matrix $H_{\mu}\left(x,p\right)$, has $r$ eigenvalues smaller than
$-C$ and that the others are greater than $C>0$. Equivalently, on
figure (b), the spectrum $\omega$ of $H_{\mu}\left(x,p\right)$ for
any $\left(x,p\right)$ is contained in the red domain.}
\end{figure}

The hypothesis (\ref{eq:gap}) is a spectral gap assumption, see figure
\ref{fig:Hypoth=00003D0000E8se}. Clearly, the model $E_{\mu}$ (\ref{eq:symbole_H_mu})
verifies this property with a gap constant $C=1$.

\subsubsection{Quantum operator}

The following definition defines a family of operators $\hat{H}_{\mu,\epsilon}=\mathrm{Op}_{\epsilon}\left(H_{\mu}\right)$
depending on a parameter $\epsilon>0$ from the symbol $H_{\mu}$
given in (\ref{eq:symbol_H_miu}). This is called quantization\footnote{In quantum mechanics, \textbf{quantization} is the procedure that
gives a quantum Hamiltonian operator $\mathrm{Op}_{\epsilon}\left(H\right)$
from a classical Hamiltonian function $H$. The parameter $\epsilon$
is sometimes written $\epsilon=h$ as the Planck constant, and the
semi classical limit is $h\rightarrow0$, where one can get some spectral
properties of $\mathrm{Op}_{\epsilon}\left(H\right)$ from the classical
dynamics generated by $H$} . 

\begin{cBoxA}{}
\begin{defn}
\cite{zworski_book_2012}The \textbf{Weyl quantization of $H_{\mu}$}
is the operator $\hat{H}_{\mu,\epsilon}=\mathrm{Op}_{\epsilon}\left(H_{\mu}\right)$
acting in $L^{2}\left(\mathbb{R}^{n};\mathbb{C}^{d}\right)$ defined
on a vector valued functions $\psi\in L^{2}\left(\mathbb{R}^{n};\mathbb{C}^{d}\right)$
by
\[
\left(\mathrm{Op}_{\epsilon}\left(H_{\mu}\right)\psi\right)\left(x\right)=\frac{1}{\left(2\pi\epsilon\right)^{n}}\int_{\mathbb{R}^{n}\times\mathbb{R}^{n}}H_{\mu}\left(\frac{x+y}{2},\xi\right)e^{i\xi\cdot\left(x-y\right)/\epsilon}\psi\left(y\right)dyd\xi
\]
\end{defn}

\end{cBoxA}

\begin{rem}
One can check that in our previous model, one has $\mathrm{Op}_{\epsilon=1}\left(E_{\mu}\right)\eq{\ref{eq:symbole_H_mu}}\hat{E}_{\mu}$
for the parameter $\epsilon=1$.
\end{rem}

\subsection{Spectral index}

In the previous subsection, we described a family of symbols $\left(H_{\mu}\right)_{\mu\in\mathbb{R}}$
and how to quantize them getting a family of operators $\left(\hat{H}_{\epsilon,\mu}\right)_{\epsilon>0,\mu\in\mathbb{R}}$.
Now, we state a theorem that defines a spectral index for the corresponding
operators.
\begin{cBoxB}{}
\begin{thm}
\cite[thm 2.2]{faure_manifestation_topol_index_2019}\label{prop:Avec-l'hypot}
A consequence of the gap assumption (\ref{eq:gap}) is that for every
$\alpha>0$ there exists $\epsilon_{0}>0$ such that for every $0<\epsilon<\epsilon_{0}$, 
\begin{itemize}
\item for any $\mu$ such that $1+\alpha<\left|\mu\right|<2$, the operator
$\hat{H}_{\mu,\epsilon}$ has \textbf{no spectrum} in the interval
$]-C+\alpha,+C-\alpha[$. 
\item for any $\mu$ such that $\left|\mu\right|\leq1+\alpha$, the operator
$\hat{H}_{\mu,\epsilon}$ has \textbf{discrete spectrum} in the interval
$]-C+\alpha,C-\alpha[$ that depends continuously on $\mu,\epsilon$. 
\end{itemize}
Consequently one can define the \textbf{spectral index} of the model
$\left(H_{\mu}\right)_{\mu}$ by
\begin{equation}
\mathcal{N}_{H}:=n_{\mathrm{in}}-n_{\mathrm{out}}\in\mathbb{Z}\label{eq:def_N_H}
\end{equation}
where the discrete spectrum has been labeled in the positive order
and $n_{\mathrm{in}}$, (respect. $n_{\mathrm{out}}$) is the label
of the the first eigenvalue below the spectral gap ``in'' (respect.
``out''). It does not depend on $\epsilon>0$.
\end{thm}

\end{cBoxB}

\begin{figure}
\begin{centering}
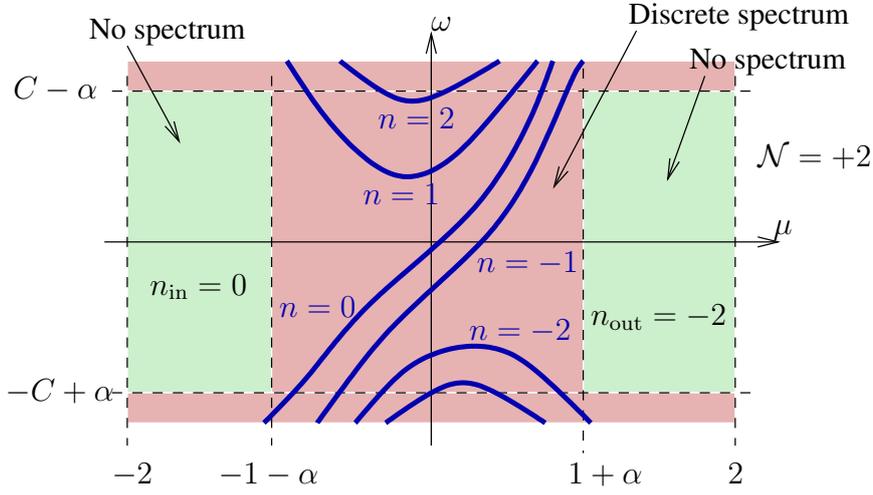
\par\end{centering}
\caption{\protect\label{fig:def_N}For $\epsilon>0$ fixed, this is a schematic
picture of the spectrum of the operator $\hat{H}_{\mu,\epsilon}$.
In this example, $\mathcal{N}=n_{\mathrm{in}}-n_{\mathrm{out}}=0-\left(-2\right)=+2$
corresponding to the fact that $\mathcal{N}=+2$ eigenvalues are moving
upward as $\mu$ increases.}
\end{figure}

See figure \ref{fig:def_N}. In other words, for a sufficiently small
$\epsilon$, the operator $\hat{H}_{\mu,\epsilon}$ presents a spectral
gap when $|\mu|>1$. Moreover, when $\mu$ goes from $-1$ to $1$,
a finite number of eigenvalues of $\hat{H}_{\mu}$ continuously crosses
the gap. The proof uses microlocal analysis, but the idea is that,
for a sufficiently small $\epsilon$, the spectrum of $\hat{H}_{\mu}$
is close to the range of $H_{\mu}$, so the red zone of figure \ref{fig:def_N}.
Moreover the compactness of the ball $\left\Vert \left(\mu,x,p\right)\right\Vert \leq1$
implies discrete spectrum for $\hat{H}_{\mu}$ in that region (this
is related to Weyl law or uncertainty principle). There are two energy
bands (lower and upper) and the spectral index counts the number of
states exchanged by them.

\begin{rem}
The integer $\mathcal{N}_{H}$ is continuous hence invariant under
continuous variations (homotopy) in the space of symbols given by
definition \ref{assump}. This means that $\mathcal{N}_{H}$ is "topological".
Then if two symbols are homotopic (one can continuously deform one
to the other) then they have the same spectral index. But the converse
is not true: we will study this point later in section \ref{sec:Topological-contact-without}. 
\end{rem}

We now give an easy but important proposition (or remark) related
to so called ``K-theory''. One can change the size of the matrix
$H_{\mu}(x,p)$ without changing the spectral index: it suffices to
add some constant eigenvalues outside of the gap:
\begin{cBoxB}{}
\begin{prop}
\label{prop:id}Consider $r\in\mathbb{N}\backslash\{0\}$ and $\omega_{0}\in\mathbb{R}$
with $|\omega_{0}|>C$ and $(H_{\mu})_{\mu}$ verifying the definition
\ref{assump}. Then $\tilde{H}_{\mu}$ defined by 
\begin{equation}
\tilde{H}_{\mu}(x,p)=H_{\mu}(x,p)\oplus\omega_{0}\mathrm{Id}_{\mathbb{C}^{r}}=\left(\begin{array}{cc}
H_{\mu}(x,p) & 0\\
0 & \omega_{0}\mathrm{Id}_{\mathbb{C}^{r}}
\end{array}\right)\label{eq:H_timlde}
\end{equation}
has the same spectral index than $H_{\mu}$:
\begin{equation}
\mathcal{N}_{\tilde{H}}=\mathcal{N}_{H\oplus\omega_{0}\mathrm{Id}_{\mathbb{C}^{r}}}=\mathcal{N}_{H}.\label{eq:N_constant}
\end{equation}
More generally we have the \textbf{additive property} for two symbols
\begin{equation}
\mathcal{N}_{H\oplus H'}=\mathcal{N}_{H}+\mathcal{N}_{H'}.\label{eq:additive_N}
\end{equation}
\end{prop}

\end{cBoxB}

\begin{proof}
The symbol $\omega_{0}\mathrm{Id}_{\mathbb{C}^{r}}$ gives a quantized
operator $\mathrm{Op}\left(\omega_{0}\mathrm{Id}_{\mathbb{C}^{r}}\right)=\omega_{0}\mathrm{Id}_{L^{2}(\mathbb{R}^{n}\otimes\mathbb{C}^{r})}$,
whose spectrum is constant eigenvalues $\{\omega_{0}\}$, independent
of $(\mu,x,p)$. Hence the spectral flow index is $\mathcal{N}_{\omega_{0}\mathrm{Id}_{\mathbb{C}^{r}}}=0$.
Due to the diagonal form of (\ref{eq:H_timlde}) the spectrum of $\mathrm{Op}\left(\tilde{H}_{\mu}\right)$
is the superposition of the spectrum of $\mathrm{Op}\left(H_{\mu}\right)$
and $\mathrm{Op}\left(\omega_{0}\mathrm{Id}_{\mathbb{C}^{r}}\right)$,
hence $\mathcal{N}_{\tilde{H}_{\mu}}=\mathcal{N}_{H}+\mathcal{N}_{\omega_{0}\mathrm{Id}_{\mathbb{C}^{r}}}=\mathcal{N}_{H}$.
\end{proof}

\subsection{Chern index}

In the previous section, we were interested in the operators $\mathrm{Op}_{\epsilon}(H_{\mu})$
in order to define the spectral index $\mathcal{N}_{H}$. Now, we
study the symbols $\left(H_{\mu}\right)_{\mu\in\mathbb{R}}$.

\begin{cBoxB}{}
\begin{prop}
\label{def:Fh}Let $(H_{\mu})_{\mu\in\mathbb{R}}$ be a family of
symbols verifying definition \ref{assump} and let
\[
S^{2n}:=\left\{ \left(\mu,x,p\right)\in\mathbb{R}^{1+2n},\quad\left\Vert \left(\mu,x,p\right)\right\Vert =1\right\} ,
\]
be the unit sphere in the space of parameters. The gap assumption
(\ref{eq:gap}) guaranties that for every parameter $\left(\mu,x,p\right)\in S^{2n}$,
we have a well defined eigenspace $F_{H}\left(\mu,x,p\right)$ of
dimension $r$ associated to the first $r$ eigenvalues $\omega_{1}\ldots\omega_{r}$
(direct sum of the first $r$ eigenspaces). This family of vector
spaces is called a \textbf{smooth complex vector bundle of rank $r$
over the sphere $S^{2n}$} and denoted $F_{H}$ or 
\[
F_{H}\rightarrow S^{2n}.
\]
The isomorphism class of this bundle $F_{H}$ denoted $\mathrm{Vect}_{\mathbb{C}}^{r}\left(S^{2n}\right)$
is characterized by the isomorphism class of the \textbf{clutching
function}{\bfseries\footnote{similarly as in definition \ref{def:-We-consider} or figure \ref{fig:Chern-index-}.
Here the equator is a sphere $S^{2n-1}$ and the clutching is a unitary
map in $\mathbb{C}^{r}$ hence an element of $U\left(r\right)$.}} on the equator $g:S^{2n-1}\to U\left(r\right)$, denoted $\pi_{2n-1}\left(U\left(r\right)\right)$
(see figure \ref{tab:Groupes-d'=00003D0000E9quivalences-de}), i.e.
\[
\mathrm{Vec}_{\mathbb{C}}^{r}(S^{2n})=\pi_{2n-1}(U(r)).
\]
\end{prop}

\end{cBoxB}

Table \ref{tab:Groupes-d'=00003D0000E9quivalences-de} shows some
class of isomorphism of vector bundles. We say that the vector bundle
$F_{H}$ is \textbf{trivial} if the clutching function is homotopic
to the identity. Remark that for $r\ge2n$, $\mathrm{Vec}_{\mathbb{C}}^{r}(S^{2n})$
is always $\mathbb{Z}$, and this is the subject of the following
important theorem.

\begin{cBoxB}{}
\begin{thm}
\label{thm:Bott-periodicity-theorem}\cite{hatcher_ktheory}\textbf{<<Bott
periodicity theorem (1959)>> }For all $n,r\in\mathbb{N}\backslash{0}$
with $r\ge n$, we have
\[
\mathrm{Vec}_{\mathbb{C}}^{r}(S^{2n})=\mathbb{Z},\qquad\mathrm{Vec}_{\mathbb{C}}^{r}(S^{2n-1})=0.
\]
\end{thm}

\end{cBoxB}

When $r<n$, the structure of $\mathrm{Vec}_{\mathbb{C}}^{r}(S^{2n})$
is more complicated and give rise to non trivial effects, see this
\href{https://en.wikipedia.org/wiki/Homotopy_groups_of_spheres\#Table}{table}.
More details will be given in the third section.
\begin{center}
\begin{table}
\begin{centering}
\begin{tabular}{|c|c|c|c|c|c|c|}
\hline 
$\mathrm{Vect}^{r}\left(S^{k}\right)$ & $S^{2}$ & $S^{3}$ & $S^{4}$ & $S^{5}$ & $S^{6}$ & $S^{7}$\tabularnewline
\hline 
\hline 
$\mathrm{Vect^{1}}$ & $\boxed{\mathbb{Z}}$ & $0$ & $0$ & $0$ & $0$ & $0$\tabularnewline
\hline 
$\mathrm{Vect^{2}}$ & $\mathbb{Z}$ & $\boxed{0}$ & $\boxed{\mathbb{Z}}$ & $\mathbb{Z}_{2}$ & $\mathbb{Z}_{2}$ & $\mathbb{Z}_{12}$\tabularnewline
\hline 
$\mathrm{Vect^{3}}$ & $\mathbb{Z}$ & $0$ & $\mathbb{Z}$ & $\boxed{0}$ & $\boxed{\mathbb{Z}}$ & $\mathbb{Z}_{6}$\tabularnewline
\hline 
$\mathrm{Vect^{4}}$ & $\mathbb{Z}$ & $0$ & $\mathbb{Z}$ & $0$ & $\mathbb{Z}$ & $\boxed{0}$\tabularnewline
\hline 
$\mathrm{Vect^{5}}$ & $\mathbb{Z}$ & $0$ & $\mathbb{Z}$ & $0$ & $\mathbb{Z}$ & $0$\tabularnewline
\hline 
 & $\vdots$ & $\vdots$ & $\vdots$ & $\vdots$ & $\vdots$ & $\vdots$\tabularnewline
\hline 
$\tilde{K}\left(S^{k}\right)$ & $\mathbb{Z}$ & $0$ & $\mathbb{Z}$ & $0$ & $\mathbb{Z}$ & $0$\tabularnewline
\hline 
\end{tabular}
\par\end{centering}
\caption{\protect\label{tab:Groupes-d'=00003D0000E9quivalences-de}Equivalence
groups of complex vector bundles of rank $r$ over sphere $S^{k}$.\foreignlanguage{french}{
$\mathrm{Vect}^{r}\left(S^{k}\right)=\pi_{k-1}\left(U\left(r\right)\right)$.}}
\end{table}
\par\end{center}

The next proposition gives a definition of the Chern index $\mathcal{C}\left(F_{H}\right)$
from the clutching function $g$ of proposition \ref{def:Fh}. This
case is significantly less straightforward than in definition \ref{def:-We-consider},
i.e. dimension $n=1$, where the Chern index simply reduces to the
winding number (or degree) of the clutching function Recall that the
\textbf{\href{https://en.wikipedia.org/wiki/Degree_of_a_continuous_mapping}{degree}}
of a map$f:S^{m}\rightarrow S^{m}$ is
\begin{equation}
\mathrm{deg}\left(f\right):=\sum_{x\in f^{-1}\left(y\right)}\mathrm{sign}\left(\mathrm{det}\left(D_{x}f\right)\right)\in\mathbb{Z},\label{eq:def_degre}
\end{equation}
which is independent of the choice of the generic point $y\in S^{m}$.
In the case $f:S^{1}\rightarrow S^{1}$, the degree $\mathrm{deg}\left(f\right)$
is also called \href{https://en.wikipedia.org/wiki/Winding_number}{winding number}
of $f$.

\begin{cBoxB}{}
\begin{prop}
\label{prop:-Let-}\cite[Section III.1.B, p.271]{booss_85} Let $g:S^{2n-1}\to U(r)$
be the clutching function of proposition \ref{def:Fh}. Suppose that
$r\ge n$. Then, we can continuously deform $g$ so that $\forall x\in S^{2n-1}$,
$g_{x}\left(e_{r}\right)=e_{r}$, where $\left(e_{1},\ldots e_{r}\right)$
is the canonical basis of $\mathbb{C}^{r}$ and $g$ restricted to
$\mathbb{C}^{r-1}\subset\mathbb{C}^{r}$ gives a function $g:S^{2n-1}\rightarrow U\left(r-1\right)$.
By iteration we get the case $r=n$ with a clutching function $g:S^{2n-1}\rightarrow U\left(n\right)$
and we define the function
\begin{equation}
f:\begin{cases}
S^{2n-1} & \rightarrow S^{2n-1}\subset\mathbb{C}^{n}\\
x & \rightarrow g_{x}\left(e_{1}\right)
\end{cases}\label{eq:def_f-1}
\end{equation}
The \textbf{Chern index} of $F_{H}$ is an integer defined by
\begin{equation}
\mathcal{C}_{H}:=\mathcal{C}(F):=\frac{\mathrm{deg}\left(f\right)}{\left(n-1\right)!}\in\mathbb{Z},\label{eq:def_C_H}
\end{equation}
where $\mathrm{deg}\left(f\right)$ is the \href{https://en.wikipedia.org/wiki/Degree_of_a_continuous_mapping}{degree}
of $f.$ Adding some constant eigenvalue in the symbol (as in (\ref{eq:H_timlde}))
does not change the Chern index:
\begin{equation}
\mathcal{C}_{H\oplus\mathrm{Id}_{\mathbb{C}^{r}}}=\mathcal{C}_{H}.\label{eq:C_constant}
\end{equation}
More generally we have the \textbf{additive property} for two symbols
\begin{equation}
\mathcal{C}_{H\oplus H'}=\mathcal{C}_{H}+\mathcal{C}_{H'}.\label{eq:additive_C}
\end{equation}

Hence if initially $r<n$, then we add a constant vector space to
each fiber $F(\mu,x,p)$ (as in (\ref{eq:H_timlde})) to reach $r=n$
and define $\mathcal{C}\left(F\right)$ using (\ref{eq:def_C_H}).
\end{prop}

\end{cBoxB}

\begin{rem}
A priori, formula (\ref{eq:def_C_H}) gives a rational number but
proposition \ref{prop:-Let-} claims that it is an integer.
\end{rem}

\subsection{Index formula}

So far, for any family of symbols $(H_{\mu})_{\mu}$ of definition
\ref{assump}, we have defined on the one hand, a spectral index $\mathcal{N}_{H}\in\mathbb{Z}$
and on the other hand, a topological index $\mathcal{C}_{H}\in\mathbb{Z}$.
In this section we will present and explain the following result.

\begin{cBoxB}{}
\begin{thm}
\textbf{\label{thm:Index-formula-Let}}\cite[thm 2.7]{faure_manifestation_topol_index_2019}\textbf{
<<Index formula>>} Let$\left(H_{\mu}\right)_{\mu}$ be a family
of symbols that satisfies the gap assumption (\ref{eq:gap}). Let
$\mathcal{N}_{H}\in\mathbb{Z}$ be the spectral index defined in (\ref{eq:def_N_H})
and let $\mathcal{C}_{H}\in\mathbb{Z}$ be the topological index defined
by (\ref{eq:def_C_H}). We have:

\begin{equation}
\mathcal{N}_{H}=\mathcal{C}_{H}.\label{eq:formule_indice}
\end{equation}
\end{thm}

\end{cBoxB}

\begin{rem}
Concerning the first model, this formula (\ref{eq:formule_indice})
has been observed in (\ref{eq:index_formula_E}).
\end{rem}

\subsection{Sketch of proof of the index formula (\ref{eq:formule_indice})}

As explained in \cite{faure_manifestation_topol_index_2019}, the
index formula (\ref{eq:formule_indice}) relies on the index theorem
on Euclidean space of Fedosov-Hörmander given in \cite[thm 7.3 p. 422]{hormander1979weyl},\cite[Thm 1, page 252]{booss_85}.
In this section we provide an convincing explanation of formula (\ref{eq:formule_indice})
using normal forms. We proceed in two steps.
\begin{enumerate}
\item \label{enu:First,-we-present}First, we present some elementary models
denoted $E^{\left(n,\mathcal{C}\right)}$, for any $n\geq1$ and $\mathcal{C}\in\mathbb{Z}$,
called normal forms, such that the phase space is $\left(x,p\right)\in\mathbb{R}^{2n}$,
the symbol is a $r\times r$ matrix with $r$ large and the model
gives the Chern index $\mathcal{C}\in\mathbb{Z}$. We then observe
that the index formula (\ref{eq:formule_indice}), holds true: 
\begin{equation}
\mathcal{N}_{E}=\mathcal{C}_{E}\label{eq:NE_CE}
\end{equation}
.
\begin{rem}
The model $E^{\left(1,1\right)}=E_{\mu}$ has already been presented
in section \ref{sec:Normal-form-model}. We present these normal form
models in detail in section \ref{subsec:The-index-formula} below.
\end{rem}

\item \label{enu:Second,-for-any}Second, for any model defined by a symbol
$H_{\mu}$ (from definition \ref{assump}), we first increase the
size $r$ of the matrix by adding constant levels (as in (\ref{eq:H_timlde})).
According to (\ref{eq:N_constant}) and (\ref{eq:C_constant}) this
operation does not change the value of $\mathcal{C}_{H}$, $\mathcal{N}_{H}$.
If $r$ is large enough, the symbol $H_{\mu}$ can be deformed continuously
to a normal form model $E^{\left(n,\mathcal{C}\right)}$ with $\mathcal{C}=\mathcal{C}_{H}$.
For this we use proposition \ref{def:Fh} and Bott periodicity theorem
\ref{thm:Bott-periodicity-theorem}. By continuity, this guaranties
that $\mathcal{N}_{H}=\mathcal{N}_{E}$. We then deduce that $\mathcal{N}_{H}=\mathcal{N}_{E}\eq{\ref{eq:NE_CE}}\mathcal{C}_{E}=\mathcal{C}=\mathcal{C}_{H}$
giving formula (\ref{eq:formule_indice}).
\end{enumerate}

\subsubsection{\protect\label{subsec:The-index-formula}The index formula with normal
forms}

We consider step \ref{enu:First,-we-present}. We consider dimension
$n\geq1$, i.e. phase space $\left(x,p\right)\in\mathbb{R}^{n}\times\mathbb{R}^{n}$
and we set 
\[
z=\left(z_{1},\ldots z_{n}\right):=x+ip\in\mathbb{C}^{n}.
\]

\begin{cBoxB}{}
\begin{prop}[Normal form model $E^{\left(n,\mathcal{C}\right)}$]
For dimension $n\geq1$, and Chern index $\mathcal{C}=\pm1$, we
define a matrix symbol of size $2^{n}$
\begin{equation}
E^{\left(n,\pm1\right)}:=\left(\begin{array}{cc}
\mp\mu\,\mathrm{Id}_{2^{n-1}} & g_{n}\left(z\right)\\
\left(g_{n}\left(z\right)\right)^{\dagger} & \pm\mu\,\mathrm{Id}_{2^{n-1}}
\end{array}\right),\label{eq:def_En_1}
\end{equation}
where the matrix $g_{n}\left(z\right)$ of size $2^{n-1}$ is defined
recurrently by
\begin{equation}
g_{n}\left(z_{1},\underbrace{z_{2},\ldots z_{n}}_{z'}\right)=\left(\begin{array}{cc}
z_{1}\,\mathrm{Id}_{2^{n-2}} & -\left(g_{n-1}\left(z'\right)\right)^{\dagger}\\
g_{n-1}\left(z'\right) & \overline{z_{1}}\,\mathrm{Id}_{2^{n-2}}
\end{array}\right),\qquad g_{1}\left(z\right)=z.\label{eq:def_g_n}
\end{equation}
\[
\]
Then for any $\mathcal{C}\in\mathbb{Z}$, we define the model $E^{\left(1,\mathcal{C}\right)}$
by direct sum of $\left|\mathcal{C}\right|$ copies
\[
\text{If \ensuremath{\mathcal{C}=0}, }E^{\left(n,\mathcal{C}\right)}:=\mathrm{Id},
\]
\[
\text{If \ensuremath{\mathcal{C}>0}, }E^{\left(n,\mathcal{C}\right)}:=E^{\left(n,1\right)}\oplus\ldots\oplus E^{\left(n,1\right)},
\]
\[
\text{If \ensuremath{\mathcal{C}<0}, }E^{\left(n,\mathcal{C}\right)}:=E^{\left(n,-1\right)}\oplus\ldots\oplus E^{\left(n,-1\right)}.
\]
For this model $E^{\left(n,\mathcal{C}\right)}$, we have
\begin{equation}
\mathcal{N}_{E}=\mathcal{C}_{E}=\mathcal{C}.\label{eq:NE_CE-1}
\end{equation}
\end{prop}

\end{cBoxB}

\begin{proof}
We proceed with different cases.
\begin{itemize}
\item The case $\mathcal{C}=0$ with a constant matrix, gives fixed eigenvectors
hence $\mathcal{C}_{E}=0$ and no exchange of states hence $\mathcal{N}_{E}=0$,
so (\ref{eq:NE_CE-1}) holds true.
\item For dimension $n=1$ and $\mathcal{C}=+1$, we get a $2\times2$ matrix
symbol $E^{\left(1,1\right)}$ depending on $\mu,z=x+ip\in\mathbb{C}$
that coincides with the first normal form model (\ref{eq:symbole_H_mu}),
indeed:
\[
E^{\left(1,1\right)}\eq{\ref{eq:def_En_1}}\left(\begin{array}{cc}
-\mu & g_{1}\left(z\right)\\
\left(g_{1}\left(z\right)\right)^{\dagger} & \pm\mu\,\mathrm{Id}_{2^{n-1}}
\end{array}\right)\eq{\ref{eq:def_g_n}}\left(\begin{array}{cc}
-\mu & x+ip\\
x-ip & \mu
\end{array}\right)\eq{\ref{eq:symbole_H_mu}}E_{\mu}.
\]
For that model, (\ref{eq:NE_CE-1}) has already been proven in (\ref{eq:index_formula_E}).
Indeed we by specific computations, we obtained $\mathcal{C}_{E^{\left(1,1\right)}}=+1$
and \emph{$\mathcal{N}_{E^{\left(1,1\right)}}=+1$}.
\item For the case $n=1$ and $\mathcal{C}=-1$, i.e. the model $E^{\left(1,-1\right)}$,
a similar computation shows that (\ref{eq:NE_CE-1}) holds true (this
is only a change of orientation on $S^{2}$).
\item More generally, for the case of any dimension $n\geq1$ , we can similarly
compute that $\mathcal{C}_{E^{\left(n,\pm1\right)}}=\pm1$ using definition
\ref{eq:def_C_H}. For this, we refer to \cite[section 1.1]{Puttmann_2003}
and \cite[prop B.25]{faure_manifestation_topol_index_2019}. We also
compute $\mathcal{N}_{E^{\left(n,\pm1\right)}}=\pm1$. For this we
first compute the Fredholm index of the operator $\hat{g}_{n}:=\mathrm{Op}_{1}\left(g_{n}\right)$
and obtain in \cite[prop B.26]{faure_manifestation_topol_index_2019}
that
\[
\mathrm{Ind}\left(\hat{g}_{n}\right):=\mathrm{dim}\mathrm{Ker}\left(\hat{g}_{n}\right)-\mathrm{dim}\mathrm{Ker}\left(\hat{g}_{n}^{\dagger}\right)=+1.
\]
Then in \cite[Lemma 2.11]{faure_manifestation_topol_index_2019} we
compute in a direct way that $\mathcal{N}_{E^{\left(n,\pm1\right)}}=\pm\mathrm{Ind}\left(\hat{g}_{n}\right)$,
from which we deduce $\mathcal{N}_{E^{\left(n,\pm1\right)}}=\pm1$.
Hence (\ref{eq:NE_CE-1}) holds true for the models $E^{\left(n,\pm1\right)}$.
\item Finally, for any $\mathcal{C}>0$ (respect. $\mathcal{C}<0$) and
any dimension $n$, we use additive properties of $\mathcal{N}$ and
$\mathcal{C}$ in (\ref{eq:additive_N}) and (\ref{eq:additive_C})
to deduce that (\ref{eq:NE_CE-1}) holds true for the models $E^{\left(n,\mathcal{C}\right)}$.
\end{itemize}
\end{proof}
\begin{rem}
In higher dimensions $n\geq1$, the iteration (\ref{eq:def_g_n})
gives more complicated normal form matrices of size $2^{n}\times2^{n}$.
For example for $n=2$, we get the symbol
\[
E^{\left(2,1\right)}\left(\mu,x,p\right)\eq{\ref{eq:def_En_1},\ref{eq:def_g_n}}\left(\begin{array}{cccc}
-\mu & 0 & z_{1} & -\overline{z_{2}}\\
0 & -\mu & z_{2} & \overline{z_{1}}\\
\overline{z_{1}} & \overline{z_{2}} & +\mu & 0\\
-z_{2} & z_{1} & 0 & +\mu
\end{array}\right),\quad\text{with }z_{1}=x_{1}+ip_{1},z_{2}=x_{2}+ip_{2}.
\]
\end{rem}

\section{\protect\label{sec:Topological-contact-without}Topological contact
without exchange}

In this last section we discuss some interesting physical phenomenon
called \textbf{``topological contact''} that may happen in the models
$\left(H_{\mu}\right)_{\mu\in\mathbb{R}}$ of definition \ref{assump},
if dimension $n$ of phase space is large enough.

Consider a model $\left(H_{\mu}\right)_{\mu\in\mathbb{R}}$ given
by definition \ref{assump}. There are two cases and subcases:
\begin{enumerate}
\item If the \textbf{vector bundle $F_{H}$ defined in proposition }\ref{def:Fh}\textbf{
is trivial} (i.e. the isomorphism class is $\left[F_{H}\right]=0$
in $\mathrm{Vec}_{\mathbb{C}}^{r}(S^{2n})$), then $\mathcal{C}=0$
and $\mathcal{N}=0$. We can perturb continuously the symbol $\left(H_{\mu}\right)_{\mu}$
toward a constant matrix $\left(\begin{array}{cc}
-1 & 0\\
0 & 1
\end{array}\right)$ (or with higher dimensions) so that we have a (big) gap for every
values of $\mu\in\mathbb{R}$, i.e. we can ``\textbf{open the gap}''.
Conversely an open gap implies that the bundle is trivial.
\item If the \textbf{vector bundle $F$ is non trivial}, (i.e. $\left[F_{H}\right]\neq0$
in $\mathrm{Vec}_{\mathbb{C}}^{r}(S^{2n})$), then it means that the
two bands are ``topologically coupled'' with a ``\textbf{topological
contact}'' and we \textbf{can not ``open the gap''}, or remove
the contact between the two bands. However there are subcases:
\begin{enumerate}
\item If $\mathcal{C}\neq0$, then $\mathcal{N}=\mathcal{C}\neq0$ also,
there are exchange states between the bands. This situations holds
if $r\geq n$ from Bott theorem \ref{thm:Bott-periodicity-theorem}.
\item If $\mathcal{C}=0$, then $\mathcal{N}=\mathcal{C}=0$ also, there
are no exchange states between the bands. From Bott theorem \ref{thm:Bott-periodicity-theorem},
this situation can not happen if $r\geq n$ but it may happen from
table \ref{tab:Groupes-d'=00003D0000E9quivalences-de}, if $r<n$
and $n\geq3$. The simplest example with $n=3$ degrees of freedom
with rank $r=2$, because $\mathrm{Vect}^{2}\left(S^{6}\right)=\mathbb{Z}_{2}=\left\{ 0,1\right\} $.
Suppose for example that $F_{H}\rightarrow S^{6}$ is non trivial
and with topological class $\left[F\right]=1\in\mathrm{Vect}^{2}\left(S^{6}\right)=\mathbb{Z}_{2}$.
It means that the two bands have a ``topological contact'', i.e.
that we can not open the gap. Nevertheless $\mathcal{N}=\mathcal{C}=0$
(because of the morphism $\left[F_{H}\right]\in\mathrm{Vect}^{2}\left(S^{6}\right)\rightarrow\mathcal{C}\in\mathbb{Z}$).
This implies that there is no exchange of states between the two bands
but there is some small gap smaller than $\sqrt{\epsilon}$, i.e.
that goes to zero in the semi classical limit $\epsilon\rightarrow0$.
We can call this a \textbf{topological contact without exchange}.
See figure below.
\end{enumerate}
\begin{center}
\begin{picture}(0,0)%
\includegraphics{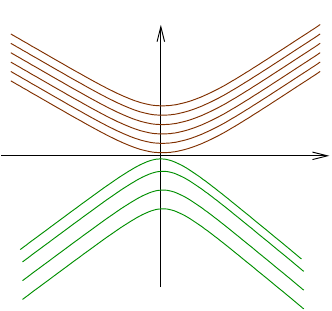}%
\end{picture}%
\setlength{\unitlength}{3947sp}%
\begin{picture}(2649,2469)(-11186,-2197)
\put(-9899, 89){\makebox(0,0)[lb]{\smash{\fontsize{12}{14.4}\usefont{T1}{ptm}{m}{n}{\color[rgb]{0,0,0}$\omega$}%
}}}
\put(-8699,-1186){\makebox(0,0)[lb]{\smash{\fontsize{12}{14.4}\usefont{T1}{ptm}{m}{n}{\color[rgb]{0,0,0}$\mu$}%
}}}
\end{picture}%

\par\end{center}

If one adds a second similar contact (at some other value of $\mu$),
then since $1+1=0$ in $\mathbb{Z}_{2}$, one recovers a trivial bundle,
the two contact annihilate themselves and one can finally ``open
the gap''. See figure below.
\begin{center}
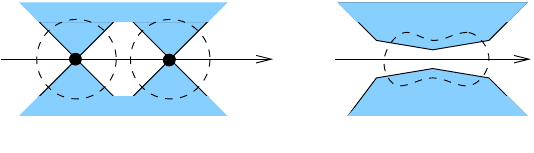
\par\end{center}

More generally these kind of phenomena may occur with models $\left(H_{\mu}\right)_{\mu}$
such that the vector bundle $F_{H}\rightarrow S^{2n}$ has rank $r<n$,
see this \href{https://en.wikipedia.org/wiki/Homotopy_groups_of_spheres\#Table_of_homotopy_groups}{table of homotopy groups}
that exhibits very rich, unexpected and complicated patterns.
\end{enumerate}
\bibliographystyle{plain}
\bibliography{/home/faure/articles/articles}

\end{document}